\documentclass{article}

\usepackage{arxiv}

\usepackage[utf8]{inputenc} % allow utf-8 input
\usepackage[T1]{fontenc}    % use 8-bit T1 fonts
\usepackage{hyperref}       % hyperlinks
\usepackage{url}            % simple URL typesetting
\usepackage{booktabs}       % professional-quality tables
\usepackage{amsfonts}       % blackboard math symbols
\usepackage{nicefrac}       % compact symbols for 1/2, etc.
\usepackage{microtype}      % microtypography
\usepackage{lipsum}
\usepackage{graphicx}

\usepackage{amsmath}
\usepackage{amssymb}

\title{Tuning Algorithms and Generators for Efficient Edge Inference}

\author{
  Rawan Naous \\
  Department of Electrical Engineering \\
  University of California Berkeley\\
  Berkeley, California 94704 \\
  \texttt{rawansn@berkeley.edu} \\
   \And
 Lazar Supic \\
  Department of Electrical Engineering \\
  University of California Berkeley\\
  Berkeley, California 94704 \\
  \texttt{lazar@berkeley.edu} \\
  \And
 Yoonhwan Kang \\
  Department of Electrical Engineering \\
  University of California Berkeley\\
  Berkeley, California 94704 \\
  \texttt{kang.yoonhwan@berkeley.edu} \\
  \And
 Ranko Sredojevic\\
  Department of Electrical Engineering \\
  University of California Berkeley\\
  Berkeley, California 94704 \\
  \texttt{rrs@berkeley.edu} \\
    \And
 Anish Singhani\\
  Department of Electrical Engineering \\
  University of California Berkeley\\
  Berkeley, California 94704 \\
  \texttt{asinghani229@berkeley.edu} \\
    \And
 Vladimir Stojanovi\c{c}\\
  Department of Electrical Engineering \\
  University of California Berkeley\\
  Berkeley, California 94704 \\
  \texttt{vlada@berkeley.edu} \\
}

\begin{document}
\maketitle

\begin{abstract}

A surge in artificial intelligence and autonomous technologies have increased the demand toward enhanced edge-processing capabilities. Computational complexity and size of state-of-the-art Deep Neural Networks (DNNs) are rising exponentially with diverse network models and larger datasets. This growth limits the performance scaling and energy-efficiency of both distributed and embedded inference platforms. Embedded designs at the edge are constrained by energy and speed limitations of available processor substrates and processor to memory communication required to fetch the model coefficients due to the inability to fit the network on a single chip. While many hardware accelerator frameworks and several network deployment frameworks have been in development, a framework is needed that allows the variety of existing architectures as well as those in development to be expressed in critical parts of the flow that perform various optimization steps. Moreover, premature architecture-blind network selection and optimization diminish the effectiveness of schedule optimizations and hardware-specific mappings that usually occur later in the flow. 
In this paper, we address these issues by creating a cross-layer software-hardware design framework that encompasses network training and model compression that is aware of and tuned to the underlying hardware architecture requirements. This approach leverages the available degrees of DNN structure and sparsity to create a converged network that can be partitioned and efficiently scheduled on the target hardware platform, minimizing data movement, and improving the overall throughput and energy.  To further streamline the design of the domain-specific hardware we leverage the high-level, flexible SoC generator platform based on RISC-V ROCC framework. This integration allows seamless extensions of the RISC-V instruction set and Chisel-based rapid generator design and design instance generation. Utilizing this approach, we implemented a silicon prototype in a 16 nm TSMC process node achieving record processing efficiency of up to 18 TOPS/W. 

\end{abstract}

\section{Introduction}

The tendency for moving towards edge computing is rising with applications getting tailored for end-users performance and privacy/security needs. The key enabler for these applications is deep neural networks (DNN) with diverse structures of networks proposed in the literature. The majority of the current networks are introducing deeper models with an increasing number of layers and parameters ~\cite{szegedy2017inception, simonyan2014very,krizhevsky2012imagenet}. Although higher levels of accuracy are attained with such approaches, they are oblivious to the underlying hardware platform and the requirements of deployment on resource-constrained edge devices~\cite{wu2019machine,fowers2018configurable}. This implementation concern has been the focus of recent research on DNN to address the memory and computational constraints with directions in algorithmic optimization such as pruning, compression and shuffling ~\cite{han2015deep,zhang2018shufflenet}, network structure and modeling with compact networks such as squeezeNet and MobileNet~\cite{iandola2016squeezenet,howard2017mobilenets} or with configurable resources such as hydraNet~\cite{teja2018hydranets} and ChamNet~\cite{dai2018chamnet}, and domain-specific accelerator design and frameworks~\cite{chen2016eyeriss, han2016eie, nvidia_l}.

Despite the vast range of proposed solutions, whether on the algorithmic or architectural levels, there is still a need for a general framework that is able to handle the diversity of the hardware/software requirements of the applications while considering the computation and real-time operation needs for embedded inference~\cite{wu2019machine}. To that end, we present a holistic approach that provides the flexibility and agility to accommodate diverse network structures and optimize computational models to adhere to the energy and performance metrics for edge computing and inference. Our solution offers a full-stack hardware-software co-design approach that allows for the incorporation of various algorithmic designs and architectural optimizations in an orchestrated manner, rather than optimizing each layer separately. Our compression algorithms, whether in terms of pruning, quantization and data structuring, are integrated within the training stages and are aware of the underlying hardware platforms hence provide an optimized reduction of the memory and communication needed for the network coefficients and enable them to fit into the on-chip memory.

Further to that, we present a complete end-to-end architecture-driven framework for code emission and scheduling on custom domain-specific accelerators on FPGA and ASIC substrates. We implement a cross-layer platform with compiler support for neural network models down to the physical tape out of a chip instance. We pave the way for flexible design space exploration and short-time to silicon by leveraging the RISC-V ROCC System-on-Chip framework.  This Chisel-based domain-specific accelerator framework covers the specific capabilities of our hardware-software co-design approach.  It primarily enables localized in-memory computing structures with tailored precision and architecture of the execution units and interconnects that are mapped to the structured compression algorithmic requirements. The generator is validated on a silicon prototype design instance implemented in 16nm TSMC process.  The chip including the RISC-V and our accelerator has an area of 6.25 $mm^2$ running at 1GHz clock speed with 20 INT4 Top/s. The compression and optimized parallel operations achieved a very low power consumption of 440 mW and 50x energy-efficiency enhancement compared to baseline unstructured pruning accelerator designs~\cite{han2016eie}.  In summary, with a full stack consideration of the design, the core of our approach is to provide end-to-end hardware/software co-design automation for deep learning inference. In particular, the novelty of our proposed approach lies in: 

\begin{itemize}

\item{\textbf{Platform aware design framework:} Introducing a cross-layer software-hardware co-design framework that is tailored to the underlying computing platform. We leverage the system-level application features to design efficient architectural flow, processing elements, interconnection network, and scheduling algorithms.}

\item{    \textbf{Novel Algorithmic Implementation:} The structured pruning algorithm we utilize uniquely identifies the decomposition characteristics for fully-connected layers. It splits the large sparse matrix into exclusive dense sub-matrices which can be processed independently with smaller processing elements (PE) and localized compact memory footprint. }

\item{    \textbf{Programmable interconnection network:} A simple scheduling algorithm is introduced to shuffle the input activations to its predetermined processing elements. The routing is achieved through a programmable multi-channel multiplexer network.}

\item{\textbf{Generic Hardware design generator:} Based on RISC-V open stack ROCC interface, the fully flexible design framework enables tuning of underlying components. The internal structure of the processing elements (PE), the number of PEs, and the interconnect infrastructure are flexible and set according to the network layer type and nature of computation. Our Focus is primarily on fully connected as an initial demonstration of capabilities, but data folding and group convolutions are applicable to split the computations over this exclusively decomposed processing model. }

\item{ \textbf{Agile hardware Design:} This approach provides wide design space exploration in a timely fashion with the feasibility of integration of different algorithms, data-precision, and architectural modifications. This full front-end and back-end flow offers a fast customizable system on chip design.}

\end{itemize}

\section{Network Model Compression}

 Looking into deep neural networks applications, there are intrinsic properties to the system that allow for error resiliency and relaxed parameter accuracy. Moreover, the redundancy in the internal connections and input nodes allows for optimization and compression for the overall system. In the algorithmic analysis of the application at hand, two bottlenecks are identified, the computational complexity and the memory hierarchy in terms of the level of communication required. Considering the memory factor, pruning has been highly effective in reducing the memory footprint, but on the expense of added pointer overhead to account for the irregularity of the sparse matrix generated~\cite{han2016eie}. Our proposed framework addresses this issue through the introduction of structured pruning. A technique that decomposes the large irregular sparse matrix into independent dense sub-matrices. Further to reducing the memory size, it also simplifies the communication and computational needs as each of the sub-matrices are processed independently.  An added feature into our framework is variable data representation. Considering the requirements of the top level application, and especially in the inference domain, the precision requirements are relaxed down to 4-bits with no loss of accuracy~\cite{dally2018hardware}.

%###################
\begin{figure}[t]
\centering
\includegraphics[width=0.7\textwidth]{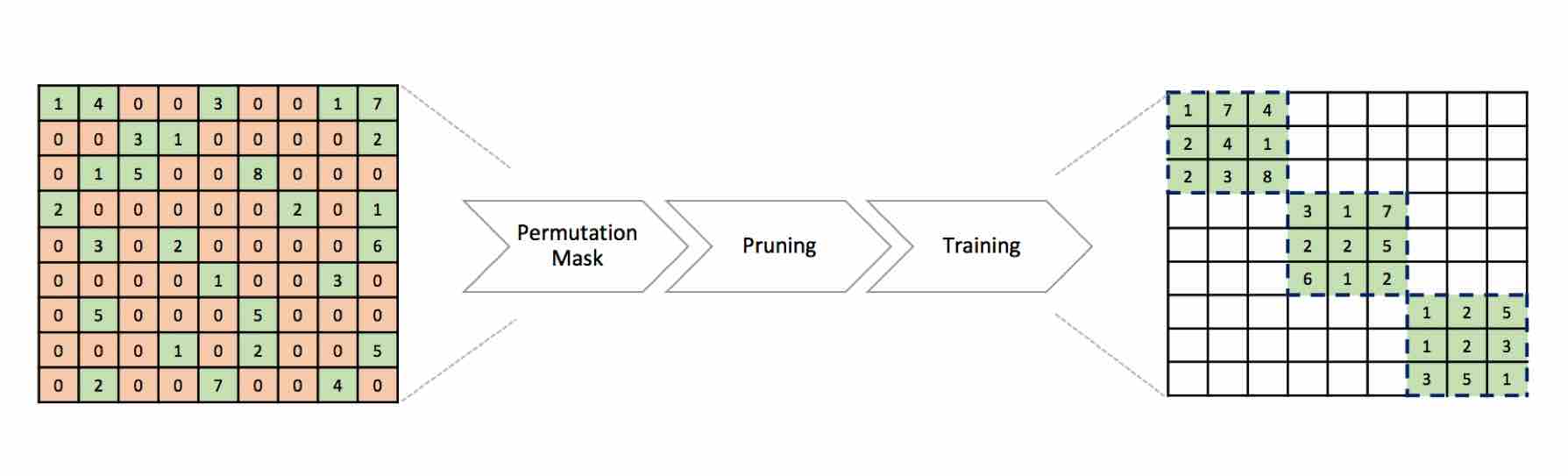}
\caption{Structured pruning with the network parameters packed into exclusive and independent blocks for efficient memory and computation.}
\label{fig:Pruning}
\end{figure}
%############################

\subsection{Structured Pruning}

Weight pruning is a technique applied to compress neural networks by exploiting the redundancy of the layer parameters. This approach reduces the size of the network layer by a large ratio reaching up to 9x - 13x, particularly for fully connected layers~\cite{han2015deep}. Nonetheless, with the pruning, the weight matrix becomes irregularly sparse since the remaining weights are spread randomly across the matrix. This sparsity limits the performance and energy-efficiency gains especially in terms of the memory access and the latency levels~\cite{supic2018mpdcompress}, with only up to 25\% of latency improvement achieved with 90\% compression level~\cite{yu2017scalpel}. These limitations are addressed in a structured pruning algorithm. The main attribute lies in molding the pruning throughout the training phase and having the non-zero weight values grow in particular allocations. With such an approach,  the layer parameters can be reordered into a set of dense matrix models that allows separable and localized computation of large matrix-vector multiplies for each network layer~\cite{sredojevic2017structured,supic2018mpdcompress}. 

Figure~\ref{fig:Pruning} shows the unstructured weight matrix of a network layer with non-zero values spread irregularly across the whole matrix.  Confining the pruning of the layer parameters to a particular distribution throughout the training phase enables an efficient block-diagonal structure outcome. The weight and activation values are permuted and packed as independent and exclusive blocks. 

The output activations for such layer will be computed separately per each block. The molding is based on having random binary masks, that are generated through random permutation of an identity matrix, point-wise multiplied with the weight matrix to ensure the block-diagonal structure.  

\begin{equation}
\overline{W}_{i} = M_{i} \circ W_{i}.
\end{equation}

Where $M-i$ is the binary mask, and $W_i$ is the current weight matrix. This pruning occurs throughout the training phase with the gradients calculated at every stage and the weights updated accordingly. The application of this compression and pruning algorithm has a negligible impact on accuracy with the degradation of less than 1\% as tested over several networks and datasets (Table 1). The accuracy loss is most severe with the aggressive pruning down to 12.5\% sparsity shown in the table. However, increasing the sparsity levels beyond that point incurs no loss of accuracy in the network performance~\cite{supic2018mpdcompress}.

Applying this technique on general purpose processors provides almost linear speedup with increasing compression rates on multi-threaded CPUs and GPUs~\cite{sredojevic2017structured}. However, the pruning is restricted to the level of parallelism and multi-threading feasible within these general purpose computing units. On the other hand, allowing the top level considerations to drive the architectural design results in a more optimized and efficient system. We verify this performance enhancement with our specialized accelerator system and elaborate on the intermediate design steps in the following sections.

\begin{table}[!h]
\label{table1}
\begin{center}
\begin{tabular}{| c | c | c |}
\hline
DNN Model & \multicolumn{2}{ c |}{Evaluation Accuracy ($\%$)}  \\ 
\cline{2-3}
& \textbf{Our algorithm} & Non-compressed \\
\hline
LeNet 300-100 & 97.3 &  98.16 \\ \hline
Deep MNIST & 99.3  & 99.3 \\ \hline
CIFAR10 & 85.2 & 86  \\ \hline
AlexNet (ImageNet) & 79.6  & 80.1  \\ \hline

\end{tabular}
\end{center}
\caption{Experimental results on several datasets and network models with 10x compression level.}
\end{table}

\subsection{Data Representation}

Closely tied to the algorithmic optimization, the appropriate number of representations for both training and inference have been extensively studied. Considering the fine tuning of the gradients and its calculation throughout the training phase, higher levels of precision are usually required to preserve the accuracy levels. On the other hand, in the inference mode, the required precision for achieving the accuracy of computation is greatly relaxed. Several models of networks are proposed with reduced precision ranging from 8 bits down to ternary~\cite{zhu2016trained} and binary representation~\cite{rastegari2016xnor}.  Moreover, the type of quantization applied has a large impact on preserving the accuracy of the system. In particular, non-uniform quantization~\cite{dally2018hardware} tends to incur no loss of accuracy for the networks operating down to 4-bit precision and even with a lower number of bits, as shown at IBM research that achieved highly accurate deep neural network models with 2 bits~\cite{choi2018bridging}. This is, in particular, the case for the reduced precision across all the network layers. More aggressive quantization levels are possible when considering the dynamic optimization of layers during training.
Accelerators as well have already provided support for reduced precision with Google TPU1~\cite{jouppi2017datacenter} operating at 8-bits and Nvidia's NVLDA~\cite{nvidia_l} open source accelerator providing support down to 4 bits.
In our work, we combine both the quantization and structured pruning iteratively during the training phase to ensure the satisfactory performance of the compressed network. Considering the hardware/software design paradigm we are proposing at the early stage, our framework is able to accommodate the integration of any quantization algorithm. Moreover, with the parameter tuning feature, different instances could be streamed out depending on the application requirements.

%###################
\begin{figure}[t]
\centering
\includegraphics[width=0.45\textwidth]{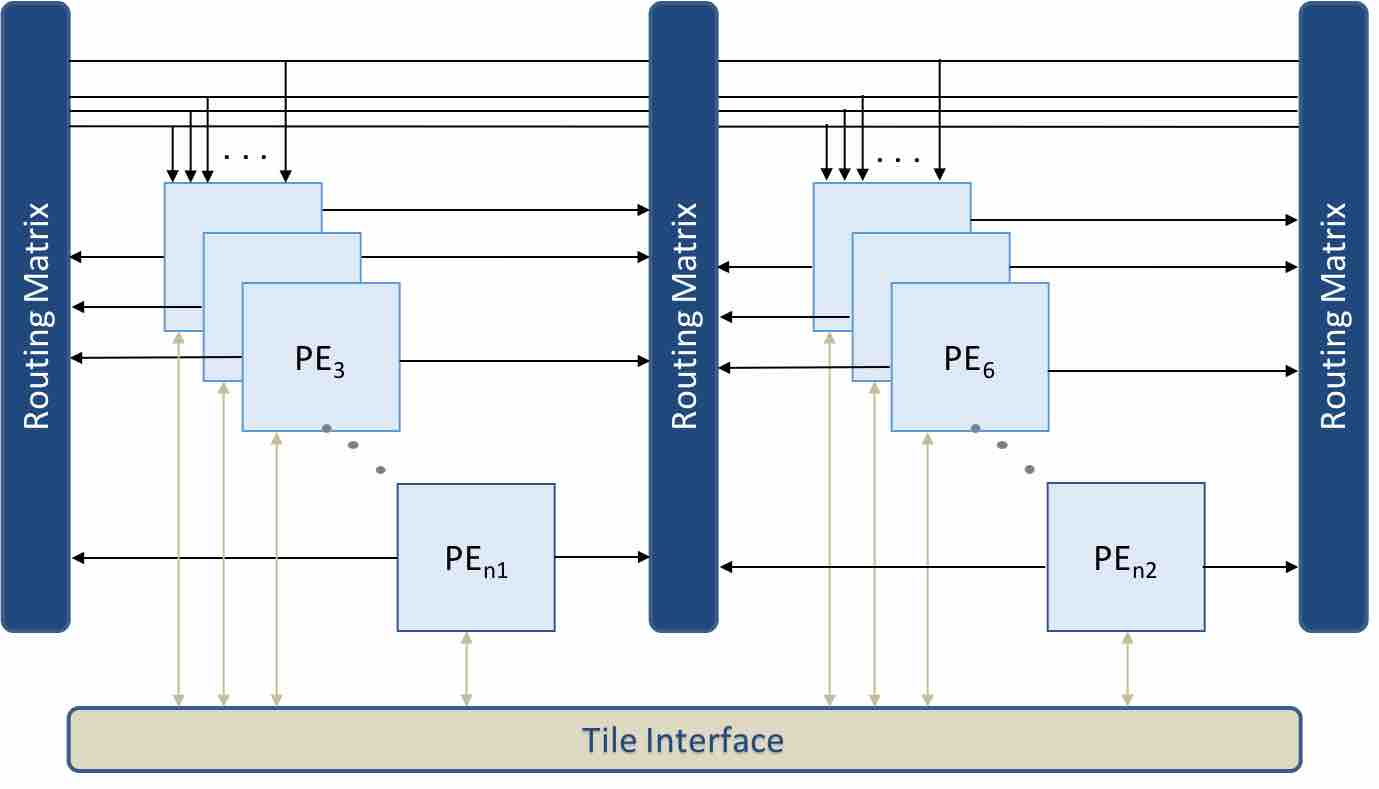}
\caption{Top level design of the accelerator engine with the permutations mapped into a flexible interconnection network connecting parallel PEs.}
\label{fig:toplevel}
\end{figure}
%############################

\section{System Architecture}

Edge and machine learning targeted designs are still in the early stages of deployment with algorithms and network models changing rapidly. This mode of operation mandates a set of features to be incorporated into the novel computing architectures.Exploiting parallelism is the main characteristic of the new workload required at the edge and paves the way for real-time processing and energy efficient designs. Additionally, the design should be agnostic to model structure and able to accommodate a diverse set of flexibility characteristics. To cover the full stack, simple programming abstraction is also an important factor to incorporate into the design as it allows an intuitive and smooth interaction with the underlying system. This section covers the details of the architectural design and the full system integration for processing neural networks.

\subsection{Accelerator Design}
Building on the structured pruning compression and the reduced data precision, the design of the underlying accelerator is based on a multi-processor array that is interconnected with an input routing matrix. Shuffling the activations between the layers is far more energy efficient,  in terms of complexity and speed, than shuffling the network weight coefficients. Figure~\ref{fig:toplevel} shows the top level design of the accelerator engine with the layers constituting of a routing matrix and array of independent processors. Each of the PEs is responsible for handling a single block within the network layer. With this split, the weight memory is local to the PE and the multiply-accumulate (MAC) operations are also exclusive within the PE. Hence, the parallel execution of the network computation is highly applicable with no interaction or communication among blocks required during layer processing. This architecture leads to an efficient hardware implementation with reduced latency and energy due to the in-block processing and memory localization.

\subsubsection{Processing Element}

In fully connected layers of neural networks, the main operation is based on a mathematical formulation of a multiply-accumulate (MAC) operation followed by a smoothing activation function.  Each of the output nodes will have an activation output value as a result of the following operation

\begin{equation}
o_i = f(\Sigma w_i * a_i + b_i)
\end{equation}

where $o_i$ is the activation value per output node, $w_i$ is the weight of the connection between the input node $i$ and the current output node,  $a_i$ is the input activation, $b_i$ is the bias value, and \textit{f} is the non-linear Rectifier Unit (ReLU). The implementation of these computations could be processed in a spatial or temporal manner.

\textbf{Temporal Processing:} The conventional mode of operation is to run parallel computations per each input activation value. Partial sums are accumulated over time for each output node. A shared register file is then used to save the partial sums as the inputs are streamed into the processing element for MAC operation. The final set of activations at the output of the network layer are all available at once with the last computation of the layer.

%###################
\begin{figure}[t]
\centering
\includegraphics[width=0.9\textwidth]{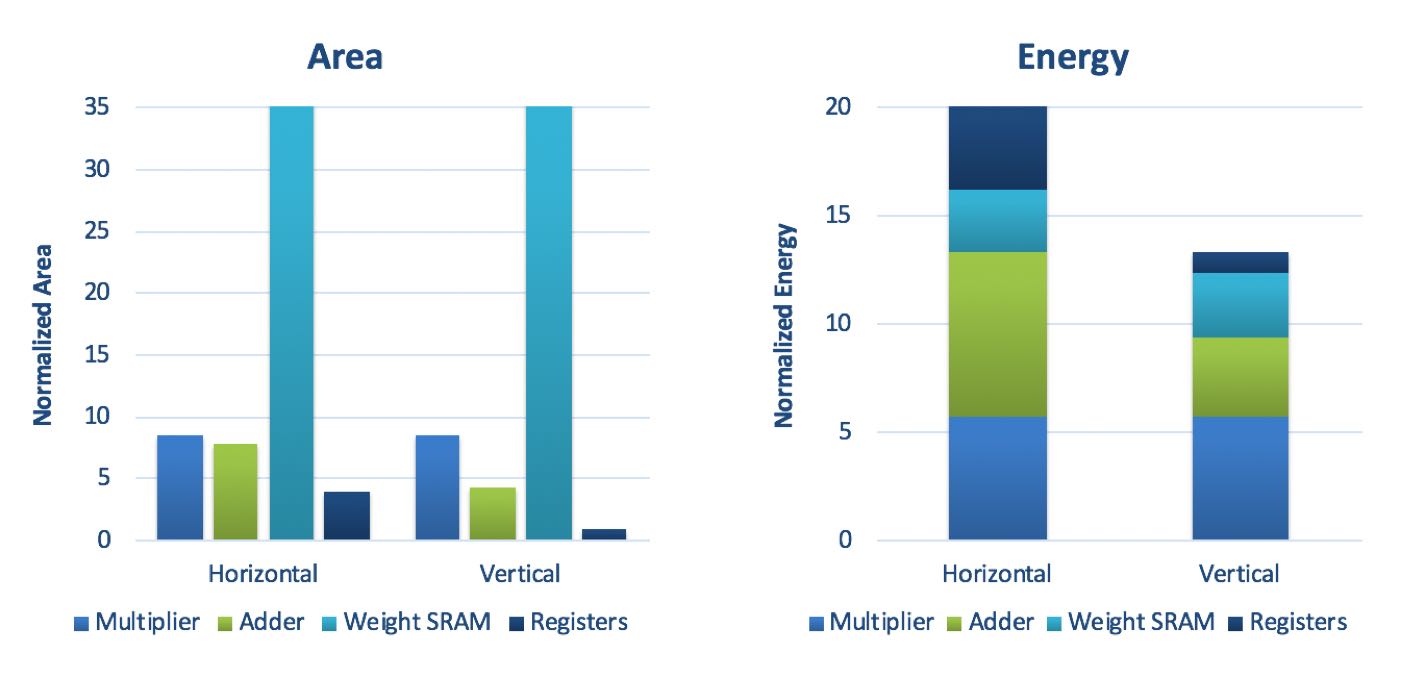}
\caption{Area and Energy analysis for the two modes of processing. Vertical processing shows a more energy-efficient implementation.}
\label{fig:alternatives_metrics}
\end{figure}
%############################

\textbf{Spatial Processing:} In an alternative approach to the temporal processing structure, computations are performed in a spatial manner. That is, per cycle a single output activation is calculated. A reduction based adder tree is used to perform the computation. To achieve this operation,  all the input activations related to a particular output value need to be available prior to the computation. The block structure and the localized operation within each block make this computational mode feasible. The input values required for a particular block are routed into the PE in a serial fashion with one activation value per cycle. Once all the activations are available in the input register, the computations are performed with a single row read from the internal weight SRAM.

Comparing the two different modes of operation, the temporal mode suffers from energy and area overheads due to the partial sum storage. A dedicated register file is required to save the partial sum values. This cost is largely reduced in the spatial mode as the output is directly calculated per cycle. The main enablers for this mode of operation are the structured pruning and quantization techniques described previously. The static scheduling makes all the activations needed available for computation, and the reduced bit precision paves the way for having a more energy and area efficient design. To quantify the performance metrics of both approaches, the energy and area for both temporal and spatial processing are calculated for a block size of 400x400 at 4-bit precision. The number of multipliers and adder tree stages are set with the block size. Hence, with our example, 400 four-bit multipliers are used. The adders are increasing in precision with the last stage having 16-bit precision. This computing scheme preserves the accuracy as the quantization is applied at the end result of the adder tree. Figure~\ref{fig:alternatives_metrics} shows the energy and area breakdown per each operation. As shown, the weight matrix and multiplication portions for the area and energy in the two approaches is the same. However, the savings are achieved in the adder tree and the register file. The adder saving is due to the incremental bit precision with every stage in the spatial reduction. Moreover, the single cycle calculation of the output activation value eliminates the need for the register file for partial sum accumulations.

Spatial processing is adopted in several accelerators such as NVDLA~\cite{nvidia_l} and DianNao~\cite{chen2014diannao}. A hybrid version of the reduction is depicted in Stitch-X accelerator~\cite{lee2018stitch} with spatial processing applied at the global level and temporal at the local computing elements. The energy efficiency of spatial processing is the main driver for its adoption. Nonetheless,  the critical path for spatial processing is more constrained due to the adder tree structure. However, with the localized computation, the level of quantization, and the parallel processing structure we are able to overcome this limitation. In our current design, we achieve a 1 GHz clock frequency and complete the processing of an output activation within 1 clock cycle. The operations include 400 four-bit multiplications, and a 9-stage adder tree with increasing precision.

%###################
\begin{figure}[t]
\centering
\includegraphics[width=0.9\textwidth]{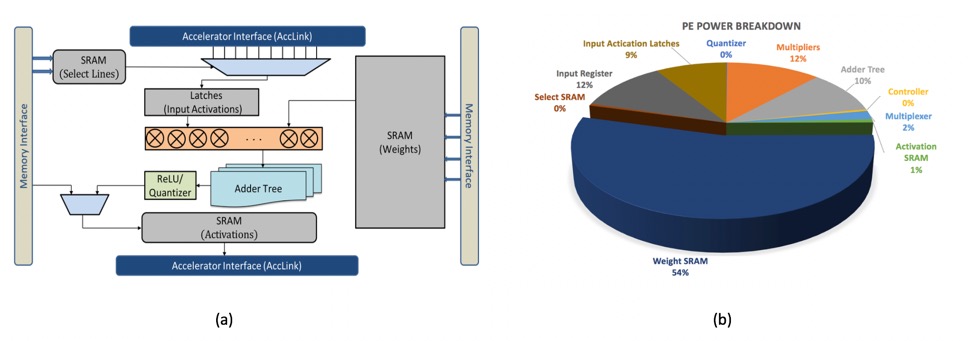}
\caption{(a)Internal structure of a single PE including the datapath flow for computation, the storage elements, and the required control. (b) Power breakdown per operation for a single PE. The values are calculated using post place\&route simulation at a 16nm technology node. The memory comprises more than 50\% of the power whereas the computation adds up to around 25\%.}
\label{fig:internal_pe}
\end{figure}
%############################

\textbf{Processing Element Structure:} With each block of the structured matrix mapped into a single PE, figure~\ref{fig:internal_pe}a shows the internal structure and flow of computations. The required modules for processing are the set of multipliers and an adder tree for computation, followed by ReLU as an activation function and a quantizer for the reduced precision operation. On the memory side, several SRAMs are needed for the storage of the layer parameters and the control operations. The weights within each block are stored in a dedicated SRAM. The input activations are saved in an input activation latch that feeds the multipliers. Once the computations are complete, the resultant output activation is saved in the output SRAM. The select SRAM is responsible for storing the static-schedule select signals for the output multiplexed crossbar.

%%%%%%%%%%%%%%%%%%%%%%%%%%%%%%%%%%%%%%%%%%%
%\begin{figure}[t]
%\centering
%\includegraphics[width=0.45\textwidth]{chip_power.png}

%\caption{Power breakdown per operation for a single PE. The values are calculated using post place\&route simulation at a 16nm technology node. The memory comprises more than 50\% of the power whereas the computation adds up to around 25\%.}
%\label{fig:power_breakdown}
%\end{figure}
%############################

Post place and route simulation results for a 16nm TSMC technology are shown in Fig.~\ref{fig:internal_pe}b. The power breakdown per each task within a cycle shows the dominance of the weight memory, with more than 50\% of the total power, and only 25\% for computation. The parameters set are for 4-bit precision and a block memory size of 400x400 weight data. Memory starts to dominate further as we increase the size of the weight SRAM. A more detailed elaboration on these design alternatives is explored in the evaluation section. The pruning algorithm sets the number of independent blocks within the layer. Then, each dense block is assigned to a dedicated PE with its own part of the data to process. Depending on the operational strategy, a compromise arises between the different performance metrics of area, power, and delay. Prioritizing fast and parallel operation sets the required memory size to account for the complete layer processing in a single step. However, such approach incurs large area and power costs and does not take into account the relaxed constraints set by the real-time operation. On the other extreme, having just a single PE with a very small weight memory size will achieve the lowest area and power measures, but imposes large latency.

%###################
\begin{figure}[t]
\centering
\includegraphics[width=0.38\textwidth, height=0.23\textheight]{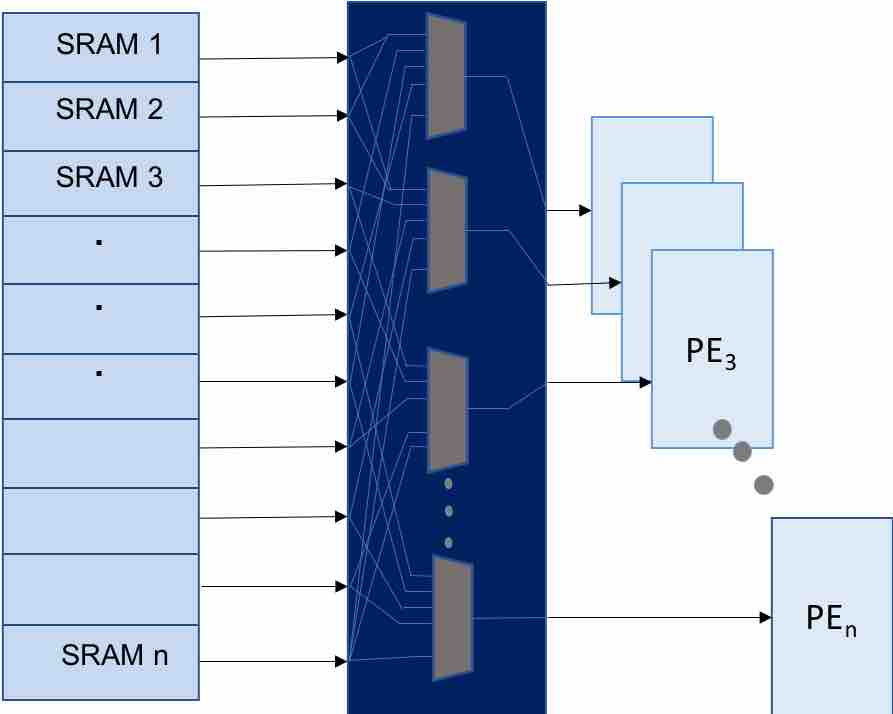}
\caption{The routing network composed of a set of multiplexers the distribute the activation values across the different PEs in the corresponding network layer. }
\label{fig:routing}
\end{figure}
%############################

%###################
\begin{figure}[t]
\centering
\includegraphics[width=0.47\textwidth, height=0.23\textheight]{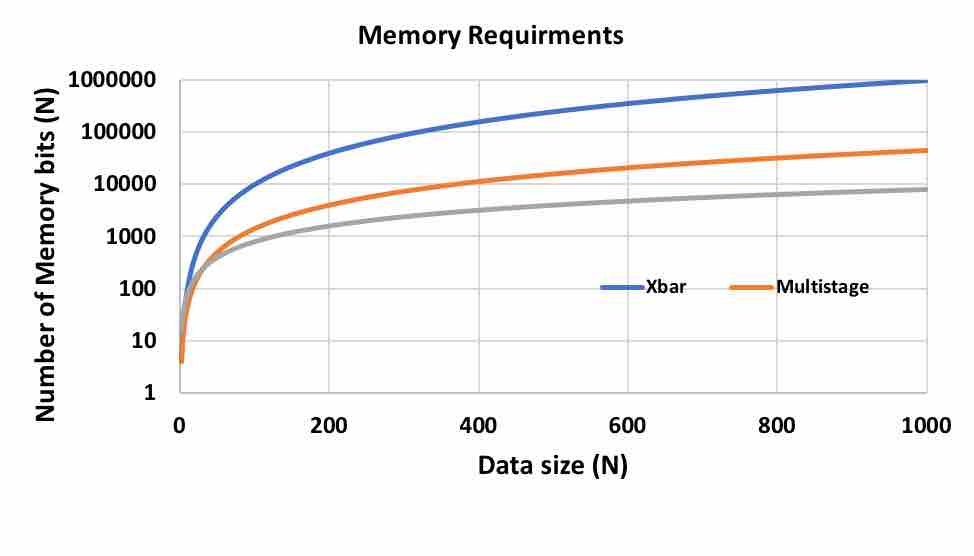}
\caption{Memory requirements per routing matrix design for Crossbar, Multistage, and current design of multiplexers.}
\label{fig:memory_routing}
\end{figure}

%############################

\subsubsection{Routing Network}
During the training phase, the structured pruning algorithm applies permutations to decompose the large weight matrix into a set of independent sub-matrices. Each of these sub-matrices is assigned to a particular processing element and is saved locally within the weight SRAM of the PE. The same permutation applies to the activation values as well. However, since the input is dynamically arriving, we use a routing network to deliver the activation values to its corresponding PE for computation. We apply a scheduling algorithm to configure the routes and the sequence of data values to be routed to the PE.  Since the permutations are known during the training phase, we can split the input into blocks, analyze the needed routes, and have a static schedule for the delivery of the input activations.

The algorithm starts with splitting the activations into blocks and sorting, in ascending order, the number of activations that need to be routed to each PE.  The block with the highest number is given the priority to choose the PE and the activation index to send over the network. In a similar fashion to round-robin implementation, the priority is moved to the next block in the sorted list. The algorithm runs with N activations routed out at every cycle to N processing elements. The scheduling is verified to avoid deadlocks and congestion in which a 1-to-1 mapping of the activation blocks and PEs are set. Each activation block sends out one activation value to a PE, and the PE will have a single activation value routed from the N blocks with no overlap. The mapping changes per cycle depending on the number of activations and the priority setting.

Figure~\ref{fig:routing} shows a block diagram of the routing matrix comprising an output-multiplexed crossbar. The output activations of a particular layer are calculated in a distributed manner over several PEs and saved internally in the output SRAM of the block. These SRAMs are considered to hold the randomly permuted input activations for the following layer. Every cycle, each of the PEs will broadcast a single activation value over the output multiplexed crossbar, which is then picked-up by the corresponding PE through setting the select value of its multiplexer. This static scheduling at compile time preserves the flexibility in applying any permutation and layer parameters during the training phase. This mode of operation saves orders of magnitude in terms of the memory requirements compared to alternative hardware-based designs. The most flexible design of routing matrices is usually a crossbar that connects each of the input to the corresponding output. However, a naive implementation would result in giant crossbar radix equal to the number of activations in a layer.  An alternative approach is to use a multistage network, such as Clos~\cite{pham2013design}, that is composed of a set of switches across each of the stages. The Clos network reduces the implementation cost but requires an optimization of the number of switches. The optimization metric is mainly deadlock avoidance, programmability, and non-blocking scheduling of the routed data. Fig.~\ref{fig:memory_routing} shows the memory size needed per interconnect design in terms of the data size (N), where (N) corresponds number of activation values to be routed across to the PEs within a single layer. As depicted, the multiplexer implementation saves one to two orders of magnitude in terms of the memory requirements compared to the multistage and crossbar structures.

\section{Accelerator Processing Unit}

Considering the co-design methodology presented earlier, we develop a flexible design generator that combines a general-purpose RISC-V processor and the parameterized accelerator through scalable on-chip fabric and instruction extension. The generator is an instance and modification of the RISC-V Rocket Core accelerator template implemented in Chisel. Additionally, we implement the complete front-end and back-end flows for a silicon prototype of an inference engine in 16nm TSMC process node. In this section, we will discuss the generator features,  compiler support,  chip details, and design space exploration covering several aspects of the overall system performance.

%###################
\begin{figure}[]
\centering
\includegraphics[width=0.45\textwidth, height=0.25\textheight]{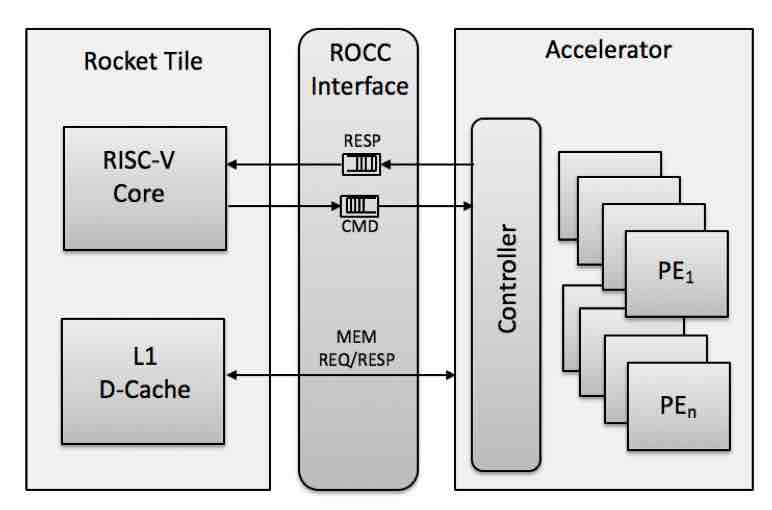}
\caption{Rocket Chip generator structure with the Rocket Core including the RISC-V and L1 cache. The RoCC interface facilitates the addition of our custom-designed accelerator with the ISA command extensions.}
\label{fig:ROCC}
\end{figure}
%############################

\subsection{Hardware Generator}

Building on our approach of prioritizing customization and flexibility in the design process, we adopt an open-source SoC generator, the Rocket Chip~\cite{asanovic2016rocket}. It encompasses a general purpose RISC-V processor with parameterized chip building libraries that makes the generator capable of producing different design instances from a high-level source. The generator is in the form of a plug-and-play environment with standardized interfaces to connect the generator libraries. The RoCC is a parametrizable interface that facilitates decoupled communication between the Rocket core and custom accelerators. Extensions to the Instruction Set Architecture (ISA) are sent over the RoCC and executed by the Core processor. As shown in Fig.~\ref{fig:ROCC}, the Rocket Tile has the RISC-V process and the L1 Cache that is directly accessible to the accelerator. Request/response commands are sent over the interface for the control and memory grant requests executed by the RISC-V processor.

Leveraging the full stack approach along with the algorithmic optimization principles, diverse innovations on the architectural side are attainable. The most prominent gains are achieved with the localized parallel execution. The first level of energy optimization and latency reduction is moving from DRAM to SRAM with around 10x savings in the energy cost~\cite{horowitz20141}. The next step is to minimize the level of communication and loading from memory as much as possible. The block structure of our computation enabled by structured pruning~\cite{sredojevic2017structured,supic2018mpdcompress} at the algorithmic level allows for having in-processor memory with minor communication among computational sub-blocks. This move paves the way for efficient and flexible mapping of the computations on the processing hardware. It also achieves around 3x of energy saving for near-processor memory operation. On top of these techniques, from the computational perspective, the large matrix operations are split into independent sub-matrices allowing for fast and parallel processing of the matrix-multiplication operation.  Additionally, architectural innovations preserve the flexibility to have different networks implemented with the varying size/number of PEs.  Hybrid integration of a general-purpose RISC-V processor along with the optimized accelerator offers the right mix between programmability and fixed functionality. These features tackle the challenges of dynamically changing neural networks in terms of model and structure.

\subsection{Software Support}
The Chisel-based generator comprises the internal structure of the processor, the interconnection network and the complete network layer. Chisel is a hardware description language, embedded in Scala, that directly generates synthesizable circuits. The functional programming features of Scala allows for the parameterization of the underlying modules and constructs with support for structured data and a high-level description of state machines and operations. The generated output is a set of low-level Verilog files that could be mapped into FPGA or ASIC designs. The functionality and RTL codes for diverse designs are generated and simulated using the cycle-accurate RTL simulator. This added feature facilitates the simulation of the entire Rocket chip instance in a significantly faster mode building on a C++ implementation. 

\begin{figure}[t]
\centering
\includegraphics[width=0.7\textwidth]{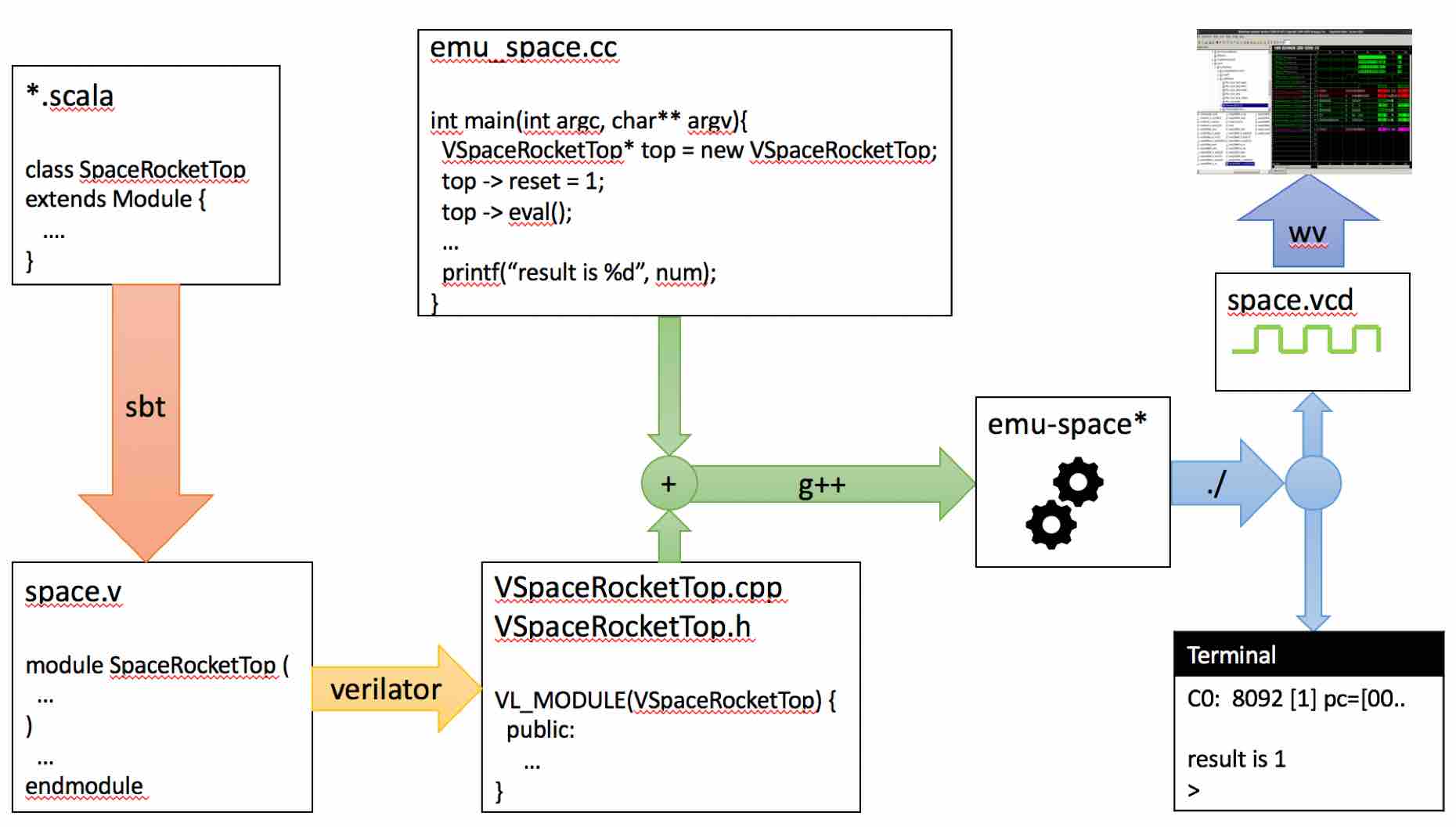}
\caption{The software support infrastructure with cycle accurate RTL simulator implemented in C++ and compiler interface for neural network processing. }
\label{fig:verification}
\end{figure}

Integrating this generator model as an end-to-end framework requires compiler support to run neural network models. In that regards, we implemented our own compiler that takes as input any high-level network model, such as tensor flow or Caffe. The first step in the conversion is parsing the model to extract the activation and weight parameters. These values are then passed into our custom compiler that translates the model into a set of Assembly code instructions to be passed into the top level accelerator as depicted in Fig.~\ref{fig:verification}. This software interface allows for easy integration and assimilation of our framework into any computing system with standard I/O or JTAG interfaces.

%###################
%############################

\subsection{Chip Design}

\begin{figure}[]
\centering
\includegraphics[width=0.9\textwidth]{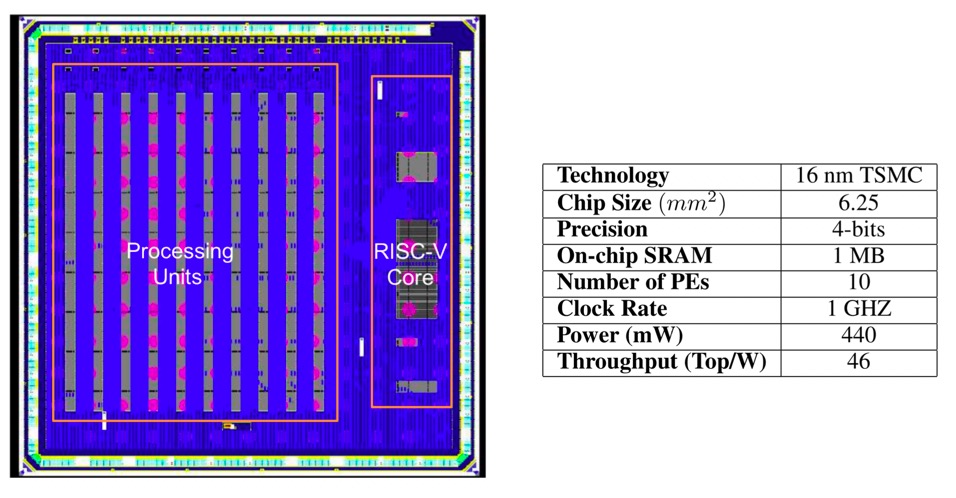}
\caption{Layout for the Accelerator Processing Unit (APU) comprising the RISC-V processor along with the array of PE.}
\label{fig:chip}
\end{figure}

We validated the hardware generator described earlier with a design instance in 16nm TSMC process. The Accelerator Processing Unit (APU) chip shown in Fig.~\ref{fig:chip} is composed of a RISC-V processor with its instruction and data caches alongside the array of processing units. The table highlights the chip specifications and the performance metrics achieved with the set parameters. This heterogeneous structure of a general-purpose processor and a dedicated accelerator preserves the flexibility to handle different network structures and models.  The design is chosen to accommodate up to 16M parameters for a fully connected (FC) layer. The accelerator side is composed of 10 PEs fully loaded with a total on-chip SRAM memory of 8 Mb. The operating frequency is 1 GHz allowing for a single layer processing capability at 400 cycles. The data precision applied is 4 bit integer for the weights and activation values with a throughput of around 16 TOPS at a power consumption of 440 mW for the overall system operation including the clock tree and the RISC-V.  The operation count includes the real multiplications,  adder tree, and the quantization steps in all the PEs.  For each cycle, there are 400 multiplications followed by 9 stages of addition in a mixed precision mode. We normalize the mixed precision operations in the adder tree down to 4-bit INT and achieve per PE around 1600 GOPs. With 10 PEs in our chip design, we achieve a total of 16 TOPS. The power measurements are reported for the operating voltage of 0.72V at fast-fast and slow-slow process corners. The throughput is calculated as the total number of operations divided by power. With the power value of 440 mV for the total operation and the 16 TOPS operation, the chip throughput achieves 36 TOPS/W. 

The APU chip is a taped out in silicon and is a standalone SoC that can accelerate the processing of sparse layers of neural networks. The convolution layers are also handled with the APU especially when framed in a matrix format~\cite{chen2016eyeriss} or addressed in a form of group convolutions. Alternatively, the chip is also compatible with other hardware platforms in which specific layers of the network could be offloaded to it for processing and speedup. Comparing with alternative accelerators,particularly models with unstructured pruning for fully connected layers~\cite{han2016eie}, the chip achieves more than 50x enhanced throughput and performance, as discussed in the related work section.

%###################
%############################

%################################
%\begin{table}[!h]
%\label{table1}
%\begin{center}
%\begin{tabular}{| l | c |}
%\hline
%\textbf{Technology} & 16 nm TSMC  \\ \hline
%\textbf{Chip Size \textbf{$(mm^2)$}} & 6.25   \\ \hline
%\textbf{Precision} & 4-bits   \\ \hline
%\textbf{On-chip SRAM} & 1 MB\\ \hline
%\textbf{Number of PEs}  &  10 \\ \hline
%\textbf{Clock Rate} & 1 GHZ   \\ \hline
%%\textbf{Frame Rate (frames/s)} & 500,000 \\ \hline
%\textbf{Power (mW)}  &  440 \\ \hline
%\textbf{Throughput (Top/W)} & 18   \\ \hline%

%\end{tabular}
%\end{center}
%\caption{Chip Specification}
%\label{table:specification}
%\end{table}

%#######################################

\subsection{Design Space Exploration}

Throughout the process of the accelerator design, several aspects need to be taken into consideration to allow for further inclusion and support for different network requirements. Primarily, in the design space exploration we focus on the memory sizes and bit-precision as they are considered the main contributors to the performance metrics.

\subsubsection{\textbf{Block Size:}} 

\begin{figure}[t]
\centering
\includegraphics[width=0.7\textwidth]{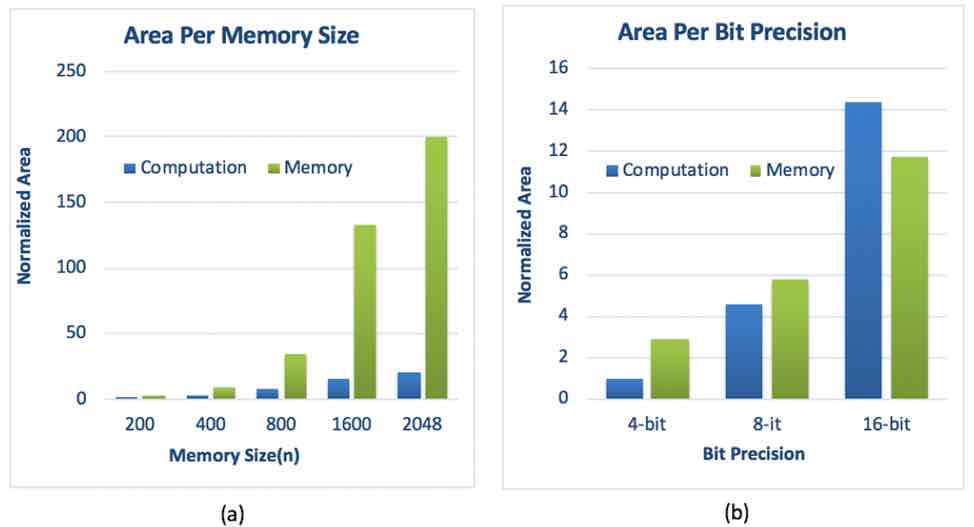}
\caption{Area performance considering different design alternatives for the PE and the full layer hardware implementation. (a) Varying memory size within the PE. (b) varying the bit precision.}
\label{fig:Design_area}
\end{figure}

\begin{figure}[t]
\centering
\includegraphics[width=0.7\textwidth]{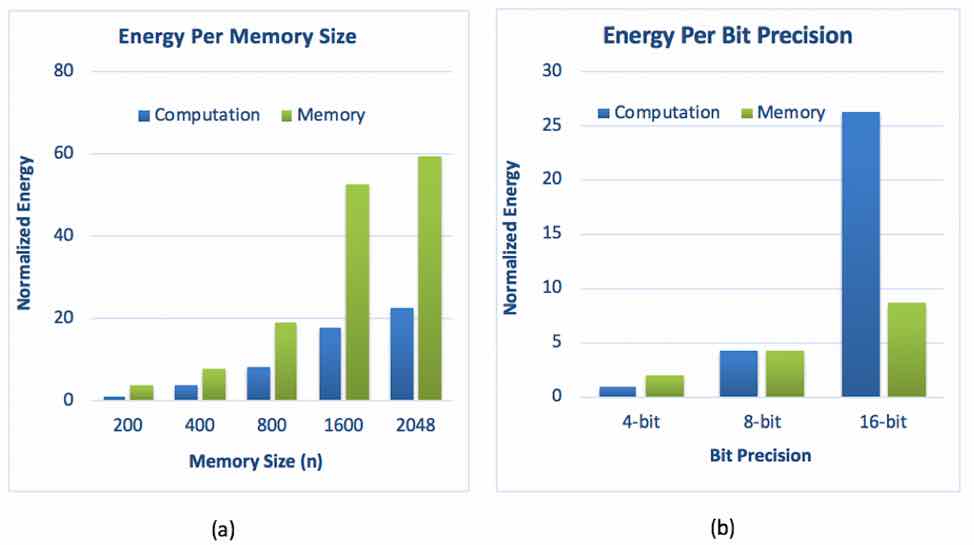}
\caption{Energy performance considering different design alternatives for the PE and the full layer hardware implementation. (a) Varying memory size within the PE. (b) varying the bit precision.}
\label{fig:Design_energy}
\end{figure}

Utilizing the functionality of the hardware generator described earlier, we generated several RTL instances of the PE with different block size, proportionally affecting the size of the internal sub-blocks. The size varied from 200 up to 2048 bits per dimension. The area and energy outcomes are reported in Figure~\ref{fig:Design_area}a and Figure~\ref{fig:Design_energy}a respectively. The area and energy for the computation scales linearly with the increasing size of the memory. This is due to the fact that number of operations that need to be executed increases with the number of entries in the memory array. On the other hand, as expected the memory incurs a quadratic increase in the area and energy values with the memory block size. Using smaller block sizes enables lower energy but complicates the routing and scheduling. The design exploration and the generator framework enable hardware tuning to a given set of applications.

%###################
%############################

%###################
%#######################

%###################
\begin{figure}[t]
\centering
\includegraphics[width=0.47\textwidth]{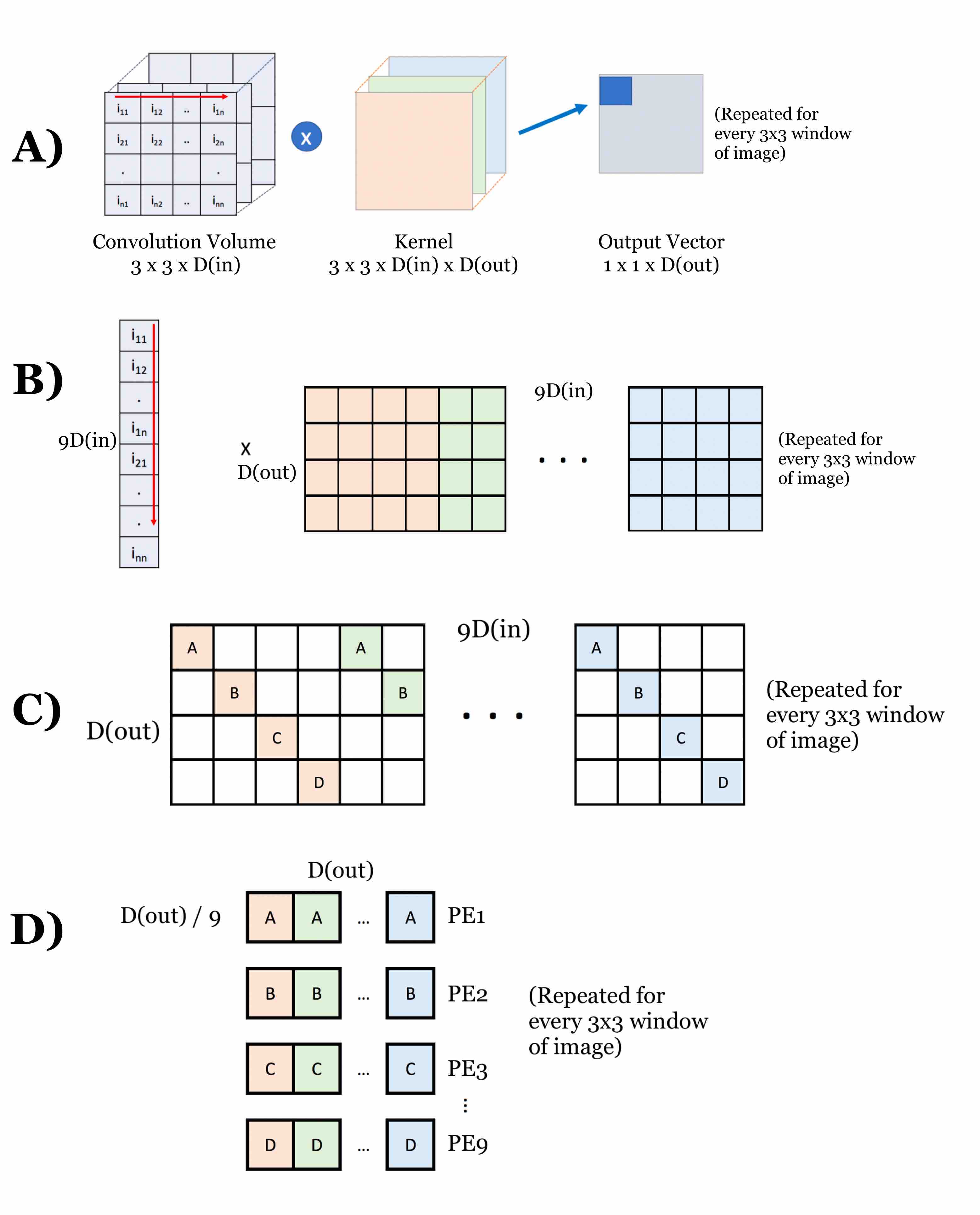}
\caption{Mapping of convolutional operations onto the PEs. (A): A standard convolution as computed when training, with 2 spatial dimensions and a depth dimension. (B): Convolution weights are unrolled along the spatial dimension (usually 3x3) as well as along the depth dimension. (C): Group convolution has convolutional weights congregated near the diagonal of the unrolled weights. (D): The structured-sparse group convolution weights are rearranged by group and distributed across the PEs.}
\label{fig:conv}
\end{figure}
%#######################

\subsubsection{\textbf{Bit Precision:}}

In this part, we consider the impact of precision on the area and energy for one PE.
The PE block size is fixed at 400x400. However, the bit precision of each entry is set differently per instance generated. We synthesized three different designs with 4, 8, and 16-bit precision respectively. Figure~\ref{fig:Design_area}b and ~\ref{fig:Design_energy}b show the area and energy dedicated for computation and memory for different bit precision. At lower bit resolution, the energy and area are both dominated by the memory. At 8-bit resolution, a break even point is achieved with almost similar performance for both functions. However, with increasing the bit precision to 16-bits, the computation starts to dominate with almost 3x more energy consumption compared to the PE local memory.

%###################
\begin{figure*}[t]
\centering
\includegraphics[width=0.92\textwidth, height=0.28\textheight]{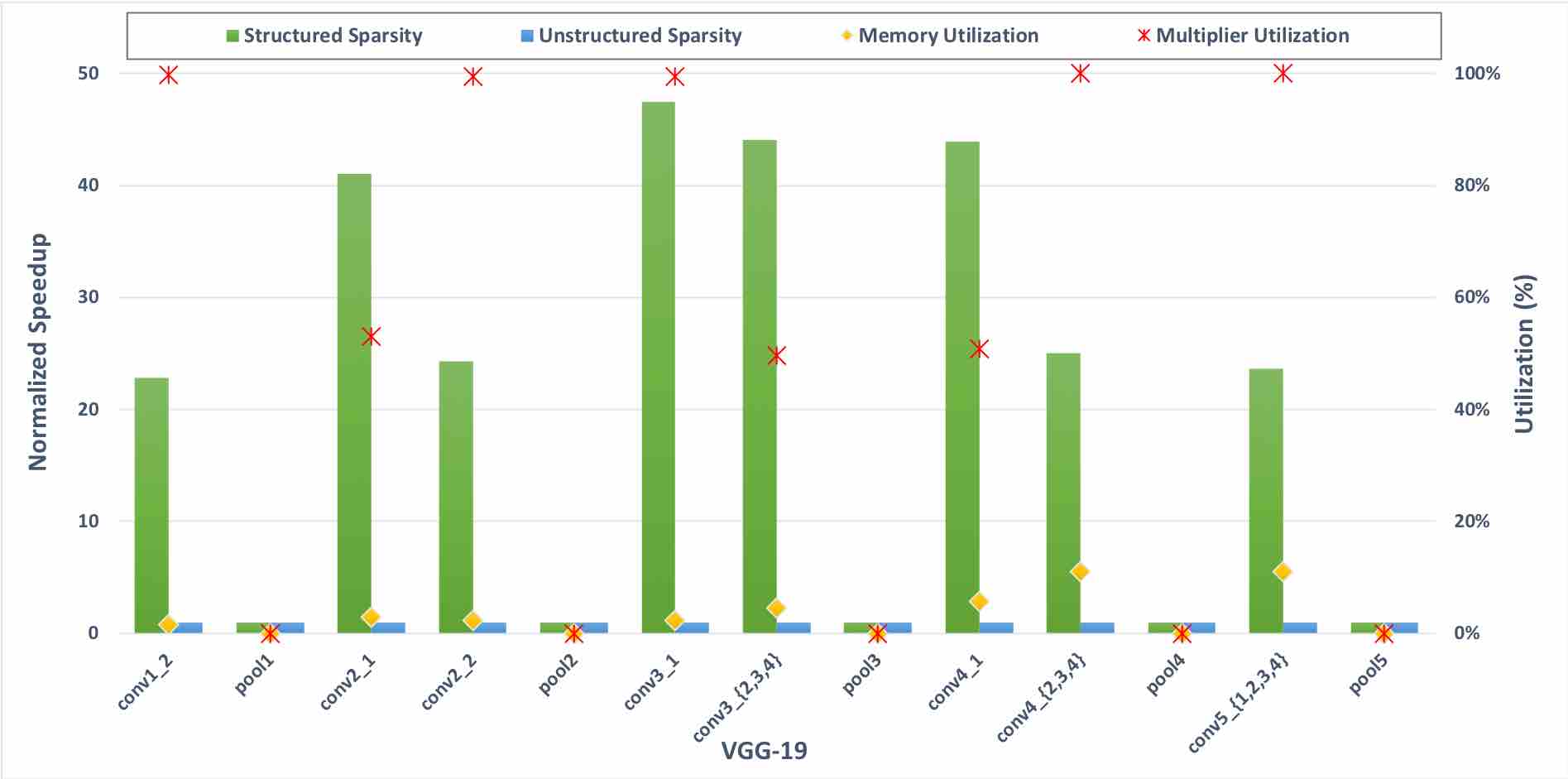}
\caption{The overall performance of the accelerator on VGG-19 with the corresponding speedup and hardware utilization}
\label{fig:VGG}
\end{figure*}
%#######################
%###################
\begin{figure*}[t]
\centering
\includegraphics[width=0.92\textwidth, height=0.28\textheight]{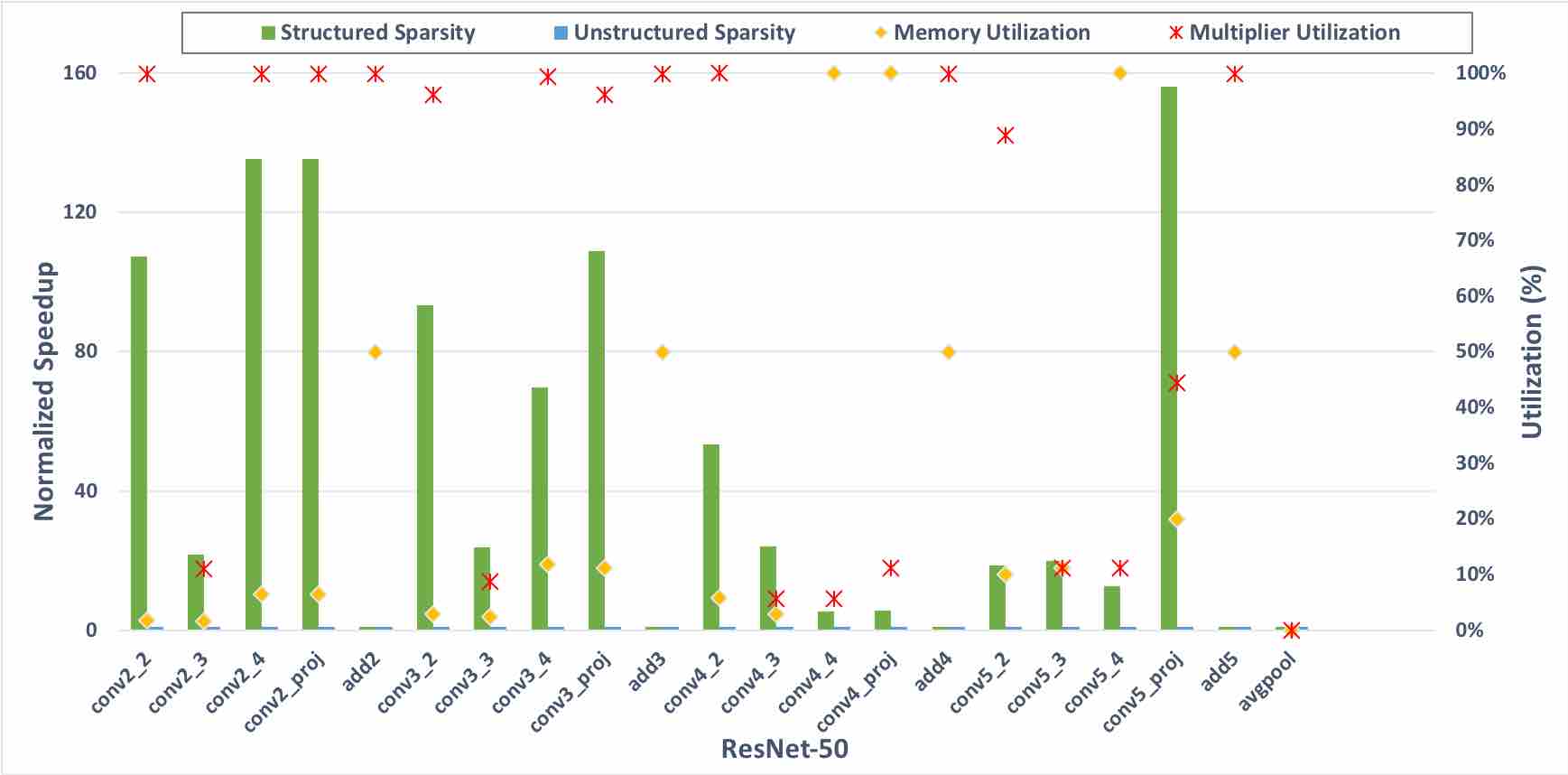}
\caption{The overall performance of the accelerator on ResNet-50 with the corresponding speedup and hardware utilization.}
\label{fig:Resnet}
\end{figure*}
%#######################

\subsubsection{Computational Layers Mapping}
\textbf{Convolution Layer:} Convolutions can be mapped onto the PEs in three ways, depending on the size and sparsity of the kernels.

I. Small non-structured-sparse kernel: If $H_k \dot W_k \dot C_{\text{in}} \leq W_{\text{PE}}$ and $C_{\text{out}} \leq H_{\text{PE}}$, then the entire convolution (for one convolution volume) can be calculated on a single PE. The remaining PEs can be used to compute other convolution volumes in parallel, reducing the number of required cycles.

II. Large non-structured-sparse kernel: If the kernel cannot be calculated on one PE, it can be distributed across multiple PEs, along the channel and/or spatial dimensions. Each portion of the kernel is multiplied by the corresponding portion of the input activations, and the pre-activation outputs are stored in memory by the RISC-V processor. The final addition and activation steps are then computed on the RISC-V and sent back to the PEs for the next layer.

III. Large structured-sparse kernel (using group convolutions): A common method of structured sparsity for convolutions is known as the group convolution. It has been used in models like AlexNet~\cite{krizhevsky2012imagenet} and ResNeXt~\cite{xie2017aggregated}, and has been shown to cause only minuscule decreases in accuracy when used as a replacement for a standard convolution~\cite{wang2019fully,zhao2018learning}. Because the group convolution has the property of structured sparsity, our accelerator can optimize networks using it very efficiently, fitting even the largest of convolutions in networks such as VGGNet (Fig.~\ref{fig:VGG}) and ResNet(Fig.~\ref{fig:Resnet}) onto just 9 513x513 PEs and computing them extremely efficiently. As depicted, the implementation of the group convolutions on the described hardware results in up to 50x speedup on VGG convolutional layers and a record 150x on ResNet. These measures are reported for a fixed number of PEs and a fixed internal structure of the PE as well. Moreover, the hardware utilization, with these fixed settings, is almost 100\% within the convolutional layers, as depicted in Fig.~\ref{fig:Resnet}, but tends to be a little low in the smaller pooling layers.

\textbf{Max Pooling: }The max-pooling layer consists primarily of comparisons and memory reading (max-pooling is a downsampling method where the maximum value in each fixed window of the image is chosen). For this reason, pooling (and other operations that do NOT consist of multiplication and addition) are run on the RISC-V core, which is flexible enough to run any operation.

\textbf{Batch Normalization:} Batch Normalization~\cite{ioffe2015batch} is a technique frequently used to accelerate neural network training. During training, batch statistics are used to update the normalization parameters after every batch. However, during inference time, the functionality of batch normalization is disabled and the approximate per-channel parameters are used instead. This means that the batch normalization is a simple linear operation. It is trivial to fold a batch normalization operation into the preceding convolution or fully-connected layer after training, and such functionality is available in libraries such as TensorFlow. 

If a nonstandard network configuration is used (such as batch normalization applied after an activation), then batch normalization can still be applied, as it is just a 1x1 convolution with a group size of 1. The normalization parameters can be converted into a convolution kernel, which can be applied to the network very efficiently due to its small group size~\cite{batch}.

\subsubsection{Natural Language Processing Networks}
Recently, neural networks based on a self-attention mechanism have become popular in the field of Natural Language Processing, as a result of their significant performance improvement over RNN-based methods \cite{vaswani2017attention, devlin2019bert, radford2019language}. In particular, \cite{vaswani2017attention} introduced a method known as Multi-Head Attention, achieving optimal results with configurations having 8-16 heads. The multi-headed attention consists of several Scaled-Dot Product attention layers (rearranged multiplication and concatenation for better parallelism):
$$\text{MultiHeadAttention}(Q, K, V) = \textstyle \sum \text{head}_i $$
$$\text{PE}_1 \longrightarrow \text{head}_1 = \text{Attention}(QW^Q_1, KW^K_1, VW^V_1) \times W^O_1 $$
$$\text{PE}_2 \longrightarrow \text{head}_2 = \text{Attention}(QW^Q_2, KW^K_2, VW^V_2) \times W^O_2 $$
$$\vdots $$
$$\text{PE}_n \longrightarrow \text{head}_n = \text{Attention}(QW^Q_n, KW^K_n, VW^V_n) \times W^O_n $$
$$\text{where Attention}(Q, K, V) = \text{softmax}(\frac{QK^T}{\sqrt{d_k}})V $$
The parallelism within the layer allows the multi-headed attention mechanism to be mapped more effectively onto the multiple-PE structure while minimizing crossover between PEs. This allows for the minimization of energy consumption when using attention-based NLP models on the APU.

\section{Related Work}

Several designs are available in the literature that consider the acceleration of neural networks~\cite{zhang2018dnnbuilder,guan2017fp, fowers2018configurable}. The FPGA accelerators have been able to achieve high levels of performance efficiency in the deployment of deep neural networks as in the case of project brainwave for cloud scale~\cite{chung2018serving}, and embedded systems~\cite{yang2019synetgy,ma2017optimizing}. Nonetheless, despite the flexibility offered in accommodating different models, the speed, area, and power efficiency attained with these accelerator models still do not reach the performance attained with specialized ASIC accelerators. Moreover, the optimization and deployment approaches vary with respect to the hardware structure and design techniques. In terms of ASIC accelerators, one of the common efforts set out lately is Nvidia~\cite{nvidia_l} open-source general purpose accelerator for Convolutional Neural Network (CNN). The main focus has been towards the optimization of the convolution layers~\cite{chen2016eyeriss, du2015shidiannao, chakradhar2010dynamically, cavigelli2017origami}.  However, when looking into the breakdown of energy within these networks, a major bottleneck lies in the large number of parameters in the fully connected layers and the cost of accessing the external memory~\cite{jouppi2017datacenter}. Although the ratio of convolutions to fully connected operations is around 20:1, the cost per operation is not the same in both. The convolutions constitute a simple multiply-accumulate (MAC) operation due to the data reuse feature available in the convolution layers for the filter weights. On the other hand, the MAC operations performed in the fully connected layers are 10x more costly as they require access to DRAM memory~\cite{sze2017hardware} to fetch the coefficients. Moreover, in convolutional layers, the coefficients are locally saved within the cache or within the processor adding 2x more efficiency in the cost. Hence, the total cost of fully connected layer operations is around 20x the cost of the convolution. Furthermore, around 60\% of the inference load in the cloud has been for processing fully connected layers ~\cite{jouppi2017datacenter}. Thus, considering the fact that the convolutions could also be transformed into a fully connected form~\cite{sze2017hardware} our accelerator is able to handle a variety of networks.

The optimization of the neural network execution has been tackled from two perspectives. First on the algorithmic side, with the compression and pruning, and second in tailoring the micro-architecture of the dedicated accelerators.

\textbf{Network Pruning} the concept of pruning the neural network and exploiting the sparsity has been explored lately either on the general purpose processors~\cite{narang2017block,sen2018sparce,yu2017scalpel,han2015deep} or dedicated accelerators. Both static pruning in which the layer weights are compressed, and dynamic pruning with zero-detection of the input activation values have been explored. In both approaches, the unstructured sparse matrix resulting from the pruning of the weights poses a limitation on the speedup and energy saving achieved from the compression technique applied due to the random memory accesses that are required. Although Scalpel~\cite{yu2017scalpel} takes into account the underlying hardware platform, it only achieves on average 1.25x speedup on GPU with Cusparse library, while structured pruning achieves 4x on the same platform \cite{sredojevic2017structured,supic2018mpdcompress}. On the other hand, considering the customized ASIC designs, EIE~\cite{han2016eie} achieves 5.12x speedup with respect to the GPU, whereas the current APU design we presented reaches up to 80x speedup for a typical fully connected layer.

\textbf{Fully-Connected Based Accelerators}
The efforts in this direction could be split into two major approaches. The first being solely focusing on the hardware aspect to optimize the computation, such as Google Tensor Processing Unit (TPU1). It is an architectural approach to computation and how to improve the access load to the memory with the implementation of systolic arrays~\cite{jouppi2017datacenter}. Despite the benefits attained with this technique, the approach considered the complete matrix operations as a huge dense matrix without leveraging the internal layer characteristics. The achieved metrics are 92 TOPS/s at 40W power. Although Google also deployed lately the Edge TPU, the focus is still on the hardware computation. The computation is set to 16-bit floating point and the power is up to 200W. Another direction is having the focus on software/algorithmic optimization. This approach has been applied in~\cite{han2016eie, reagen2016minerva} where a specific accelerator engine is designed to implement a set of system level algorithms of threshold pruning, coding, and quantization. However, when measuring the benefits attained, there is a mismatch between the theoretical gains and the actual achieved results. For example, a compression of 90\% lead to only 25\% speedup in  the operation~\cite{yu2017scalpel}, mostly due to the random memory access patterns. Hence, further architectural considerations needs to be taken into account such as the sparse matrix operations and the efficiency of computation with such approach.

\begin{figure}[t]
\centering
\includegraphics[width=0.45\textwidth]{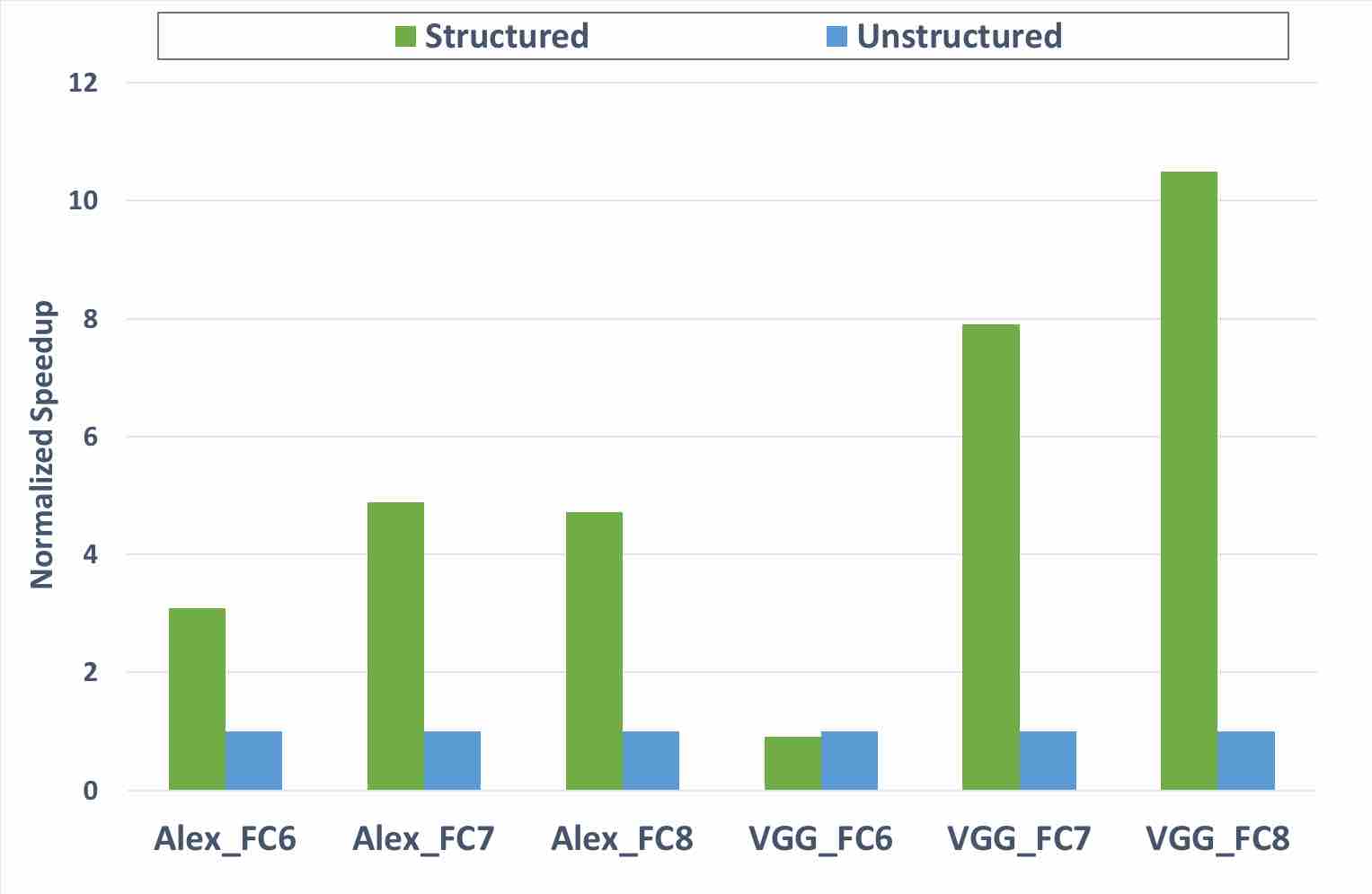}
\caption{The performance enhancement of the structured pruning architecture in comparison to unstructured pruning models.}
\label{fig:fc_compu}
\end{figure}

We compare the APU metrics with the accelerator that target fully connected layers with unstructured pruning~\cite{han2016eie} to show the improvement attained with the adoption of this architectural approach. Fig.~\ref{fig:fc_compu} shows the speedup attained with the structured pruning approach on fully connected layers vs the unstructured pruning. The comparison is based on having a fixed size of 512x512 memory and 9 PEs to compute each layer. As depicted, up to 10x speedup is attained with the our approach, taking into consideration that the structured pruning only accounts for the weight sparsity and does not incorporate the activation sparsity as in the case of~\cite{han2016eie}. In regards to the VGGFC6, the full layer does not fit completely into the set number of PEs, so folding of operations is required, hence the decrease in the speedup. However, with a larger number of PEs, a speedup of at least 2x is achieved. Moreover, with the data being equally distributed among the PEs, the load is almost always balanced and leads to efficient deployment and utilization of the available hardware. 

%###################
%#######################

\section{Conclusion}
In this paper, we present an agile design framework for deep neural network inference accelerators. A novel Hardware-Software Co-design methodology is introduced with an algorithmic driven hardware generator capable of tailoring and automated design exploration adapted to the top level application. A heterogeneous design is implemented in which a general purpose RISC-V processor is integrated with our specific accelerator through a generic interface and instruction set extensions. The structure preserves the flexibility and generality to incorporate different neural network models and sizes. 

A multi-processor system-on-chip accelerator instance is generated building primarily on algorithmic optimization and architectural innovations. In-processor memory along with localized parallel processing achieve record performance metrics for the inference accelerator. A holistic benchmarking of the operation lead to low power, high-speed, energy and area efficient embedded system with 36 TOPS/W. Although tailored for the inference applications on edge and embedded devices, this accelerator framework scales well to other datacenter-based large scale inference models that necessitate model parallel distributed computation~ \cite{chung2018serving}.

\bibliographystyle{IEEEtran}
\bibliography{paper}

\begin{thebibliography}{10}
\providecommand{\url}[1]{#1}
\csname url@samestyle\endcsname
\providecommand{\newblock}{\relax}
\providecommand{\bibinfo}[2]{#2}
\providecommand{\BIBentrySTDinterwordspacing}{\spaceskip=0pt\relax}
\providecommand{\BIBentryALTinterwordstretchfactor}{4}
\providecommand{\BIBentryALTinterwordspacing}{\spaceskip=\fontdimen2\font plus
\BIBentryALTinterwordstretchfactor\fontdimen3\font minus
  \fontdimen4\font\relax}
\providecommand{\BIBforeignlanguage}[2]{{%
\expandafter\ifx\csname l@#1\endcsname\relax
\typeout{** WARNING: IEEEtran.bst: No hyphenation pattern has been}%
\typeout{** loaded for the language `#1'. Using the pattern for}%
\typeout{** the default language instead.}%
\else
\language=\csname l@#1\endcsname
\fi
#2}}
\providecommand{\BIBdecl}{\relax}
\BIBdecl

\bibitem{szegedy2017inception}
C.~Szegedy, S.~Ioffe, V.~Vanhoucke, and A.~A. Alemi, ``Inception-v4,
  inception-resnet and the impact of residual connections on learning.'' in
  \emph{AAAI}, vol.~4, 2017, p.~12.

\bibitem{simonyan2014very}
K.~Simonyan and A.~Zisserman, ``Very deep convolutional networks for
  large-scale image recognition,'' \emph{arXiv preprint arXiv:1409.1556}, 2014.

\bibitem{krizhevsky2012imagenet}
A.~Krizhevsky, I.~Sutskever, and G.~E. Hinton, ``Imagenet classification with
  deep convolutional neural networks,'' in \emph{Advances in neural information
  processing systems}, 2012, pp. 1097--1105.

\bibitem{wu2019machine}
C.-J. Wu, D.~Brooks, K.~Chen, D.~Chen, S.~Choudhury, M.~Dukhan, K.~Hazelwood,
  E.~Isaac, Y.~Jia, B.~Jia \emph{et~al.}, ``Machine learning at facebook:
  Understanding inference at the edge,'' in \emph{2019 IEEE International
  Symposium on High Performance Computer Architecture (HPCA)}.\hskip 1em plus
  0.5em minus 0.4em\relax IEEE, 2019, pp. 331--344.

\bibitem{fowers2018configurable}
J.~Fowers, K.~Ovtcharov, M.~Papamichael, T.~Massengill, M.~Liu, D.~Lo,
  S.~Alkalay, M.~Haselman, L.~Adams, M.~Ghandi \emph{et~al.}, ``A configurable
  cloud-scale dnn processor for real-time ai,'' in \emph{Proceedings of the
  45th Annual International Symposium on Computer Architecture}.\hskip 1em plus
  0.5em minus 0.4em\relax IEEE Press, 2018, pp. 1--14.

\bibitem{han2015deep}
S.~Han, H.~Mao, and W.~J. Dally, ``Deep compression: Compressing deep neural
  networks with pruning, trained quantization and huffman coding,'' \emph{arXiv
  preprint arXiv:1510.00149}, 2015.

\bibitem{zhang2018shufflenet}
X.~Zhang, X.~Zhou, M.~Lin, and J.~Sun, ``Shufflenet: An extremely efficient
  convolutional neural network for mobile devices,'' in \emph{Proceedings of
  the IEEE Conference on Computer Vision and Pattern Recognition}, 2018, pp.
  6848--6856.

\bibitem{iandola2016squeezenet}
F.~N. Iandola, S.~Han, M.~W. Moskewicz, K.~Ashraf, W.~J. Dally, and K.~Keutzer,
  ``Squeezenet: Alexnet-level accuracy with 50x fewer parameters and< 0.5 mb
  model size,'' \emph{arXiv preprint arXiv:1602.07360}, 2016.

\bibitem{howard2017mobilenets}
A.~G. Howard, M.~Zhu, B.~Chen, D.~Kalenichenko, W.~Wang, T.~Weyand,
  M.~Andreetto, and H.~Adam, ``Mobilenets: Efficient convolutional neural
  networks for mobile vision applications,'' \emph{arXiv preprint
  arXiv:1704.04861}, 2017.

\bibitem{teja2018hydranets}
R.~Teja~Mullapudi, W.~R. Mark, N.~Shazeer, and K.~Fatahalian, ``Hydranets:
  Specialized dynamic architectures for efficient inference,'' in
  \emph{Proceedings of the IEEE Conference on Computer Vision and Pattern
  Recognition}, 2018, pp. 8080--8089.

\bibitem{dai2018chamnet}
X.~Dai, P.~Zhang, B.~Wu, H.~Yin, F.~Sun, Y.~Wang, M.~Dukhan, Y.~Hu, Y.~Wu,
  Y.~Jia \emph{et~al.}, ``Chamnet: Towards efficient network design through
  platform-aware model adaptation,'' \emph{arXiv preprint arXiv:1812.08934},
  2018.

\bibitem{chen2016eyeriss}
Y.-H. Chen, J.~Emer, and V.~Sze, ``Eyeriss: A spatial architecture for
  energy-efficient dataflow for convolutional neural networks,'' in \emph{ACM
  SIGARCH Computer Architecture News}, vol.~44, no.~3.\hskip 1em plus 0.5em
  minus 0.4em\relax IEEE Press, 2016, pp. 367--379.

\bibitem{han2016eie}
S.~Han, X.~Liu, H.~Mao, J.~Pu, A.~Pedram, M.~A. Horowitz, and W.~J. Dally,
  ``Eie: efficient inference engine on compressed deep neural network,'' in
  \emph{Computer Architecture (ISCA), 2016 ACM/IEEE 43rd Annual International
  Symposium on}.\hskip 1em plus 0.5em minus 0.4em\relax IEEE, 2016, pp.
  243--254.

\bibitem{nvidia_l}
\BIBentryALTinterwordspacing
NVIDIA, ``Nvidia deep learning accelerator (nvdla),'' 2017. [Online].
  Available: \url{https://github.com/nvdla/}
\BIBentrySTDinterwordspacing

\bibitem{dally2018hardware}
W.~J. Dally, C.~T. Gray, J.~Poulton, B.~Khailany, J.~Wilson, and L.~Dennison,
  ``Hardware-enabled artificial intelligence,'' in \emph{2018 IEEE Symposium on
  VLSI Circuits}.\hskip 1em plus 0.5em minus 0.4em\relax IEEE, 2018, pp. 3--6.

\bibitem{supic2018mpdcompress}
L.~Supic, R.~Naous, R.~Sredojevic, A.~Faust, and V.~Stojanovic,
  ``Mpdcompress-matrix permutation decomposition algorithm for deep neural
  network compression,'' \emph{arXiv preprint arXiv:1805.12085}, 2018.

\bibitem{yu2017scalpel}
J.~Yu, A.~Lukefahr, D.~Palframan, G.~Dasika, R.~Das, and S.~Mahlke, ``Scalpel:
  Customizing dnn pruning to the underlying hardware parallelism,'' in
  \emph{ACM SIGARCH Computer Architecture News}, vol.~45, no.~2.\hskip 1em plus
  0.5em minus 0.4em\relax ACM, 2017, pp. 548--560.

\bibitem{sredojevic2017structured}
R.~Sredojevic, S.~Cheng, L.~Supic, R.~Naous, and V.~Stojanovic, ``Structured
  deep neural network pruning via matrix pivoting,'' \emph{arXiv preprint
  arXiv:1712.01084}, 2017.

\bibitem{zhu2016trained}
C.~Zhu, S.~Han, H.~Mao, and W.~J. Dally, ``Trained ternary quantization,''
  \emph{arXiv preprint arXiv:1612.01064}, 2016.

\bibitem{rastegari2016xnor}
M.~Rastegari, V.~Ordonez, J.~Redmon, and A.~Farhadi, ``Xnor-net: Imagenet
  classification using binary convolutional neural networks,'' in
  \emph{European Conference on Computer Vision}.\hskip 1em plus 0.5em minus
  0.4em\relax Springer, 2016, pp. 525--542.

\bibitem{choi2018bridging}
J.~Choi, P.~I.-J. Chuang, Z.~Wang, S.~Venkataramani, V.~Srinivasan, and
  K.~Gopalakrishnan, ``Bridging the accuracy gap for 2-bit quantized neural
  networks (qnn),'' \emph{arXiv preprint arXiv:1807.06964}, 2018.

\bibitem{jouppi2017datacenter}
N.~P. Jouppi, C.~Young, N.~Patil, D.~Patterson, G.~Agrawal, R.~Bajwa, S.~Bates,
  S.~Bhatia, N.~Boden, A.~Borchers \emph{et~al.}, ``In-datacenter performance
  analysis of a tensor processing unit,'' in \emph{Computer Architecture
  (ISCA), 2017 ACM/IEEE 44th Annual International Symposium on}.\hskip 1em plus
  0.5em minus 0.4em\relax IEEE, 2017, pp. 1--12.

\bibitem{chen2014diannao}
T.~Chen, Z.~Du, N.~Sun, J.~Wang, C.~Wu, Y.~Chen, and O.~Temam, ``Diannao: A
  small-footprint high-throughput accelerator for ubiquitous
  machine-learning,'' in \emph{ACM Sigplan Notices}, vol.~49, no.~4.\hskip 1em
  plus 0.5em minus 0.4em\relax ACM, 2014, pp. 269--284.

\bibitem{lee2018stitch}
C.-E. Lee, Y.~S. Shao, J.-F. Zhang, A.~Parashar, J.~Emer, S.~W. Keckler, and
  Z.~Zhang, ``Stitch-x: An accelerator architecture for exploiting unstructured
  sparsity in deep neural networks,'' in \emph{SysML}, 2018.

\bibitem{pham2013design}
P.-H. Pham, J.~Song, J.~Park, and C.~Kim, ``Design and implementation of an
  on-chip permutation network for multiprocessor system-on-chip,'' \emph{IEEE
  transactions on very large scale integration (VLSI) systems}, vol.~21, no.~1,
  pp. 173--177, 2013.

\bibitem{asanovic2016rocket}
K.~Asanovic, R.~Avizienis, J.~Bachrach, S.~Beamer, D.~Biancolin, C.~Celio,
  H.~Cook, D.~Dabbelt, J.~Hauser, A.~Izraelevitz \emph{et~al.}, ``The rocket
  chip generator,'' \emph{EECS Department, University of California, Berkeley,
  Tech. Rep. UCB/EECS-2016-17}, 2016.

\bibitem{horowitz20141}
M.~Horowitz, ``1.1 computing's energy problem (and what we can do about it),''
  in \emph{Solid-State Circuits Conference Digest of Technical Papers (ISSCC),
  2014 IEEE International}.\hskip 1em plus 0.5em minus 0.4em\relax IEEE, 2014,
  pp. 10--14.

\bibitem{xie2017aggregated}
S.~Xie, R.~Girshick, P.~Doll{\'a}r, Z.~Tu, and K.~He, ``Aggregated residual
  transformations for deep neural networks,'' in \emph{Proceedings of the IEEE
  conference on computer vision and pattern recognition}, 2017, pp. 1492--1500.

\bibitem{wang2019fully}
X.~Wang, M.~Kan, S.~Shan, and X.~Chen, ``Fully learnable group convolution for
  acceleration of deep neural networks,'' in \emph{Proceedings of the IEEE
  Conference on Computer Vision and Pattern Recognition}, 2019, pp. 9049--9058.

\bibitem{zhao2018learning}
R.~Zhao and W.~Luk, ``Learning grouped convolution for efficient domain
  adaptation,'' \emph{arXiv preprint arXiv:1811.09341}, 2018.

\bibitem{ioffe2015batch}
S.~Ioffe and C.~Szegedy, ``Batch normalization: Accelerating deep network
  training by reducing internal covariate shift,'' \emph{arXiv preprint
  arXiv:1502.03167}, 2015.

\bibitem{batch}
\BIBentryALTinterwordspacing
tehnokv, ``Fusing batch normalization and convolution in runtime.'' [Online].
  Available: \url{https://tehnokv.com/posts/fusing-batchnorm-and-conv/}
\BIBentrySTDinterwordspacing

\bibitem{vaswani2017attention}
A.~Vaswani, N.~Shazeer, N.~Parmar, J.~Uszkoreit, L.~Jones, A.~N. Gomez, L.~u.
  Kaiser, and I.~Polosukhin, ``Attention is all you need,'' in \emph{Advances
  in Neural Information Processing Systems 30}, I.~Guyon, U.~V. Luxburg,
  S.~Bengio, H.~Wallach, R.~Fergus, S.~Vishwanathan, and R.~Garnett, Eds.,
  2017, pp. 5998--6008.

\bibitem{devlin2019bert}
J.~Devlin, M.-W. Chang, K.~Lee, and K.~Toutanova, ``{BERT}: Pre-training of
  deep bidirectional transformers for language understanding,'' in
  \emph{Proceedings of the 2019 Conference of the North {A}merican Chapter of
  the Association for Computational Linguistics: Human Language Technologies,
  Volume 1 (Long and Short Papers)}.\hskip 1em plus 0.5em minus 0.4em\relax
  Association for Computational Linguistics, 2019, pp. 4171--4186.

\bibitem{radford2019language}
A.~Radford, J.~Wu, R.~Child, D.~Luan, D.~Amodei, and I.~Sutskever, ``Language
  models are unsupervised multitask learners,'' 2019.

\bibitem{zhang2018dnnbuilder}
X.~Zhang, J.~Wang, C.~Zhu, Y.~Lin, J.~Xiong, W.-m. Hwu, and D.~Chen,
  ``Dnnbuilder: an automated tool for building high-performance dnn hardware
  accelerators for fpgas,'' in \emph{Proceedings of the International
  Conference on Computer-Aided Design}.\hskip 1em plus 0.5em minus 0.4em\relax
  ACM, 2018, p.~56.

\bibitem{guan2017fp}
Y.~Guan, H.~Liang, N.~Xu, W.~Wang, S.~Shi, X.~Chen, G.~Sun, W.~Zhang, and
  J.~Cong, ``Fp-dnn: An automated framework for mapping deep neural networks
  onto fpgas with rtl-hls hybrid templates,'' in \emph{2017 IEEE 25th Annual
  International Symposium on Field-Programmable Custom Computing Machines
  (FCCM)}.\hskip 1em plus 0.5em minus 0.4em\relax IEEE, 2017, pp. 152--159.

\bibitem{chung2018serving}
E.~Chung, J.~Fowers, K.~Ovtcharov, M.~Papamichael, A.~Caulfield, T.~Massengill,
  M.~Liu, D.~Lo, S.~Alkalay, M.~Haselman \emph{et~al.}, ``Serving dnns in real
  time at datacenter scale with project brainwave,'' \emph{IEEE Micro},
  vol.~38, no.~2, pp. 8--20, 2018.

\bibitem{yang2019synetgy}
Y.~Yang, Q.~Huang, B.~Wu, T.~Zhang, L.~Ma, G.~Gambardella, M.~Blott,
  L.~Lavagno, K.~Vissers, J.~Wawrzynek \emph{et~al.}, ``Synetgy:
  Algorithm-hardware co-design for convnet accelerators on embedded fpgas,'' in
  \emph{Proceedings of the 2019 ACM/SIGDA International Symposium on
  Field-Programmable Gate Arrays}.\hskip 1em plus 0.5em minus 0.4em\relax ACM,
  2019, pp. 23--32.

\bibitem{ma2017optimizing}
Y.~Ma, Y.~Cao, S.~Vrudhula, and J.-s. Seo, ``Optimizing loop operation and
  dataflow in fpga acceleration of deep convolutional neural networks,'' in
  \emph{Proceedings of the 2017 ACM/SIGDA International Symposium on
  Field-Programmable Gate Arrays}.\hskip 1em plus 0.5em minus 0.4em\relax ACM,
  2017, pp. 45--54.

\bibitem{du2015shidiannao}
Z.~Du, R.~Fasthuber, T.~Chen, P.~Ienne, L.~Li, T.~Luo, X.~Feng, Y.~Chen, and
  O.~Temam, ``Shidiannao: Shifting vision processing closer to the sensor,'' in
  \emph{ACM SIGARCH Computer Architecture News}, vol.~43, no.~3.\hskip 1em plus
  0.5em minus 0.4em\relax ACM, 2015, pp. 92--104.

\bibitem{chakradhar2010dynamically}
S.~Chakradhar, M.~Sankaradas, V.~Jakkula, and S.~Cadambi, ``A dynamically
  configurable coprocessor for convolutional neural networks,'' \emph{ACM
  SIGARCH Computer Architecture News}, vol.~38, no.~3, pp. 247--257, 2010.

\bibitem{cavigelli2017origami}
L.~Cavigelli and L.~Benini, ``Origami: A 803-gop/s/w convolutional network
  accelerator,'' \emph{IEEE Transactions on Circuits and Systems for Video
  Technology}, vol.~27, no.~11, pp. 2461--2475, 2017.

\bibitem{sze2017hardware}
V.~Sze, Y.-H. Chen, J.~Emer, A.~Suleiman, and Z.~Zhang, ``Hardware for machine
  learning: Challenges and opportunities,'' in \emph{2017 IEEE Custom
  Integrated Circuits Conference (CICC)}.\hskip 1em plus 0.5em minus
  0.4em\relax IEEE, 2017, pp. 1--8.

\bibitem{narang2017block}
S.~Narang, E.~Undersander, and G.~Diamos, ``Block-sparse recurrent neural
  networks,'' \emph{arXiv preprint arXiv:1711.02782}, 2017.

\bibitem{sen2018sparce}
S.~Sen, S.~Jain, S.~Venkataramani, and A.~Raghunathan, ``Sparce: Sparsity aware
  general purpose core extensions to accelerate deep neural networks,''
  \emph{IEEE Transactions on Computers}, 2018.

\bibitem{reagen2016minerva}
B.~Reagen, P.~Whatmough, R.~Adolf, S.~Rama, H.~Lee, S.~K. Lee, J.~M.
  Hern{\'a}ndez-Lobato, G.-Y. Wei, and D.~Brooks, ``Minerva: Enabling
  low-power, highly-accurate deep neural network accelerators,'' in \emph{ACM
  SIGARCH Computer Architecture News}, vol.~44, no.~3.\hskip 1em plus 0.5em
  minus 0.4em\relax IEEE Press, 2016, pp. 267--278.

\end{thebibliography}
%%%%%%%%%%%%%%%%%%%%%%%%%%%%%%%%%%%%

\end{document}